\documentclass[sigconf]{acmart}

\usepackage{amsmath}
\usepackage{xcolor,colortbl}

\usepackage{graphicx}
\usepackage{here}
\PassOptionsToPackage{hyphens}{url}
\usepackage{url} 
\usepackage{textcomp} 
\usepackage{multicol}
\usepackage{subcaption}
\usepackage{multirow}
\usepackage{xspace}

\usepackage{multirow}
\usepackage{acmart-taps}

\AtBeginEnvironment{quote}{\itshape}
\AtBeginDocument{%
\providecommand\BibTeX{{%
\normalfont B\kern-0.5em{\scshape i\kern-0.25em b}\kern-0.8em\TeX}}}

\aptLtoXcmd{}{\renewenvironment{quote}[1][0.04\linewidth]
  {\list{}{\leftmargin=#1\rightmargin=#1}\item\relax}{\endlist}
}

\usepackage{color}
\definecolor{orange}{RGB}{255,127,0}
\definecolor{limegreen}{RGB}{50, 205, 50}
\definecolor{violet}{RGB}{148,0,211}

\newcommand{\camera}[1]{\textcolor{black}{#1}}

\makeatletter
\def\Hline{
  \noalign{\ifnum0=`}\fi\hrule \@height 4.\arrayrulewidth \futurelet
   \reserved@a\@xhline}
\makeatother

\AtBeginDocument{%
  \providecommand\BibTeX{{%
    Bib\TeX}}}

\copyrightyear{2025}
\acmYear{2025}
\setcopyright{cc}
\setcctype{by}
\acmConference[UIST '25]{The 38th Annual ACM Symposium on User Interface Software and Technology}{September 28-October 1, 2025}{Busan, Republic of Korea}
\acmBooktitle{The 38th Annual ACM Symposium on User Interface Software and Technology (UIST '25), September 28-October 1, 2025, Busan, Republic of Korea}
\acmDOI{10.1145/3746059.3747697}
\acmISBN{979-8-4007-2037-6/2025/09}

\acmSubmissionID{8955}

\newcommand{\system}{\texttt{NarraGuide}}

\def\oursys{\texttt{NarraGuide}}

\begin{document}

\title{NarraGuide: an LLM-based Narrative Mobile Robot for Remote Place Exploration}

\author{Yaxin Hu}
\orcid{0000-0003-4462-0140}
\authornote{Both authors contributed equally to this research.}
\affiliation{%
  \institution{Department of Computer Sciences\\University of Wisconsin--Madison}
  \streetaddress{Department of Computer Sciences, University of Wisconsin--Madison}
  \country{} 
}
\email{yaxin.hu@wisc.edu}

\author{Arissa J. Sato}
\orcid{0000-0002-1103-8050}
\authornotemark[1]
\affiliation{%
  \institution{Department of Computer Sciences\\University of Wisconsin--Madison}
  \streetaddress{Department of Computer Sciences, University of Wisconsin--Madison}
  \country{} 
}
\email{asato@wisc.edu}

\author{Jingxin Du}
\orcid{0009-0007-7017-3213}
\affiliation{%
  \institution{Department of Computer Sciences\\University of Wisconsin--Madison}
  \streetaddress{Department of Computer Sciences, University of Wisconsin--Madison}
  \country{} 
}
\email{djx@cs.wisc.edu}

\author{Chenming Ye}
\orcid{0009-0007-0371-3101}
\affiliation{%
  \institution{Department of Computer Sciences\\University of Wisconsin--Madison}
  \streetaddress{Department of Computer Sciences, University of Wisconsin--Madison}
  \country{} 
}
\email{cye52@wisc.edu}

\author{Anjun Zhu}
\orcid{0009-0001-9583-8900}
\affiliation{%
  \institution{Department of Computer Sciences\\University of Wisconsin--Madison}
  \streetaddress{Department of Computer Sciences, University of Wisconsin--Madison}
  \country{} 
}
\email{azhu39@wisc.edu}

\author{Pragathi Praveena}
\orcid{0000-0002-8696-6265}
\affiliation{%
  \institution{The Robotics Institute\\Carnegie Mellon University}
  \streetaddress{Carnegie Mellon University}
  \country{} 
}
\email{pragathi@cmu.edu}

\author{Bilge Mutlu}
\orcid{0000-0002-9456-1495}
\affiliation{%
  \institution{Department of Computer Sciences\\University of Wisconsin--Madison}
  \streetaddress{Department of Computer Sciences, University of Wisconsin--Madison}
  \country{} 
}
\email{bilge@cs.wisc.edu}

\renewcommand{\shortauthors}{Hu et al.}

\begin{abstract}
Robotic telepresence enables users to navigate and experience remote environments. However, effective navigation and situational awareness depend on users' prior knowledge of the environment, limiting the usefulness of these systems for exploring unfamiliar places. We explore how integrating location-aware LLM-based narrative capabilities into a mobile robot can support remote exploration. We developed a prototype system, called \system{}, that provides narrative guidance for users to explore and learn about a remote place through a dialogue-based interface. We deployed our prototype in a geology museum, where remote participants ($n=20$) used the robot to tour the museum. Our findings reveal how users perceived the robot's role, engaged in dialogue in the tour, and expressed preferences for bystander encountering. Our work demonstrates the potential of LLM-enabled robotic capabilities to deliver location-aware narrative guidance and enrich the experience of exploring remote environments. 

\end{abstract}



\begin{teaserfigure}
    \includegraphics[width=\textwidth]{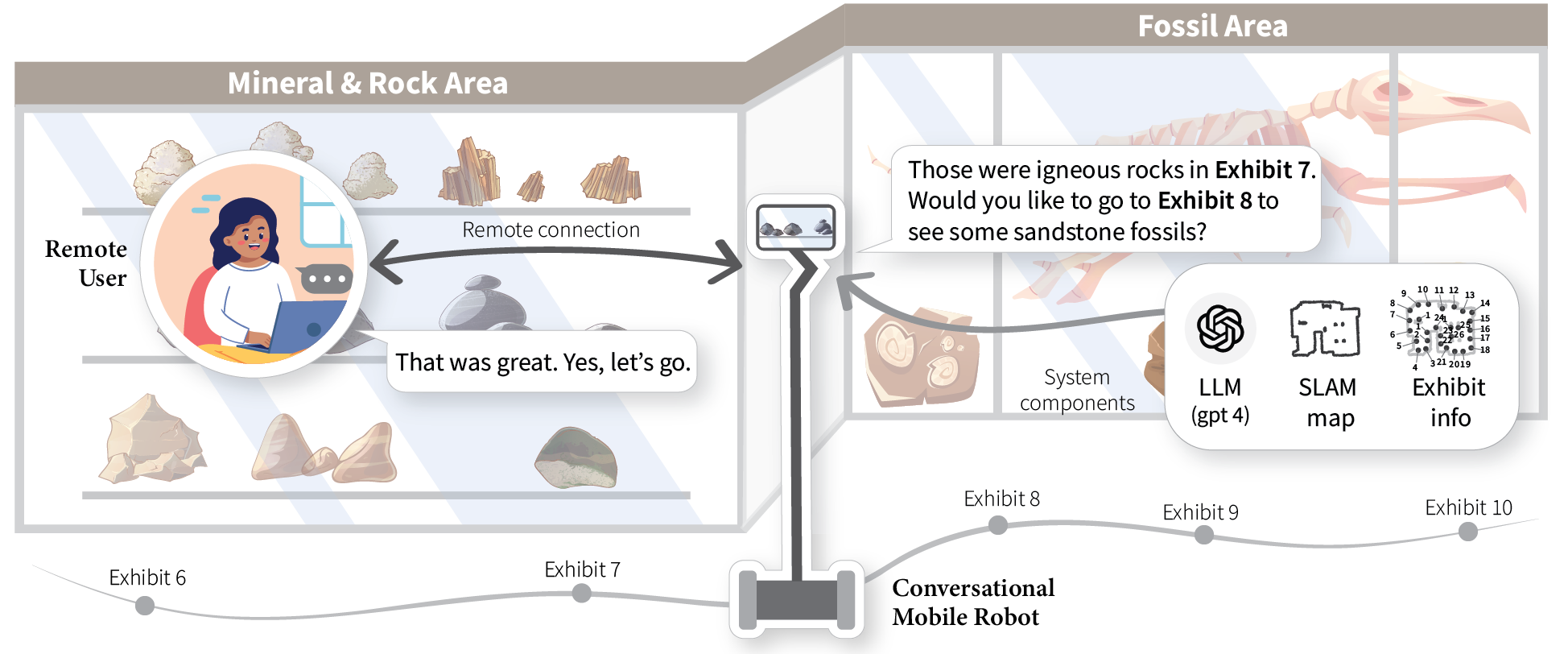}
    \caption{In this paper, we present \system{}, an \textit{LLM-based narrative mobile robot} that guides users to explore a remote \camera{physical place in real-time} and uses dialogue to deliver rich location-based information to enhance the experience.}
    \Description{figure description}
    \label{fig: teaser}
\end{teaserfigure}

\maketitle

\section{Introduction} \label{sec: intro}
 
Remote technologies enable people to connect with places they cannot visit in person for a variety of reasons, including disability, travel costs, and lack of access. Previous work has studied the use of remote technologies such as telepresence robots, 360-degree cameras, and mixed reality, opening up opportunities for users to visit remote cultural heritage sites \cite{karreman2015visiting, cantucci2022autonomous}, explore urban areas \cite{heshmat2018geocaching} and experience natural habitats \cite{tota2022telepresence}. During these experiences, technologies enable users to visually access the remote place and navigate the environment. 

Most current remote technologies require users to control navigation and focus on viewing the remote environment. However, teleoperation has been found challenging \cite{kristoffersson2013review}, and effective navigation, situation awareness, and experiencing the environment require users to have prior knowledge of the environment. Large Language Models (LLMs) have the potential to address these challenges toward augmenting the remote experience. In this work, we built an LLM-based narrative mobile robot that guides remote users around a remote location and provides location-aware dialogue to enrich the experience. The scenario below illustrates our vision for how a narrative mobile robot can help a user explore a park remotely. 

\begin{quote}
\textbf{User persona}: Sandy wants to see fall foliage in the park where she used to live. After moving to a new city, she can rarely visit her old neighborhood, and the city she currently lives in has a very short fall season. She wishes to experience fall foliage in her old neighborhood again via \system{}.
\end{quote}

\begin{quote}
\textbf{Robot}: What is your favorite spot in this park? 

\textbf{User}: I enjoyed sitting on the benches under the big oak trees.  

\textbf{Robot}: Oak trees are beautiful in the fall. There is one oak tree turning red right over there. Shall we move over there to check it out first? 

\textbf{User}: Sounds good, take me there.

\textbf{Robot}: We're on our way to the oak tree... Did you know that oak trees have an average lifespan of 150--250 years and can grow over 1,000 years old? 
\end{quote}

In this scenario, the robot guides the user to an area of interest in the remote place and engages in conversation with the user about aspects of the area. The robot provides rich narratives contextualized and based on the objects nearby in the remote place featuring factual, historical, or cultural information.

In this paper, we explored the design space of LLM-enabled remote robotic technology by creating \system---a reference design that allowed users to tour a museum. We investigated the following two research questions: (1) \textit{how can we support self-guided remote exploration with a location-aware narrative agent integrated into a robot}; (2) \textit{what use patterns emerge through user interactions with this robot?} We designed and prototyped a narrative mobile robot \camera{in a geology museum that autonomously guides the remote user to explore the museum in real-time and provide rich narratives capturing the historical and cultural backgrounds of the exhibits.}

\system{} utilizes two features to support remote place exploration: (1) autonomous robot navigation enabled by simultaneous localization and mapping (SLAM) and (2) location-aware conversational capability powered by an LLM. \camera{We encoded the museum's tour guidebook into the SLAM map, which the LLM utilized to generate the narrations and stories of each exhibit.} In the museum, a user can give the robot commands to move around, or alternatively, the robot can guide the user through autonomous navigation. \camera{The robot can passively respond to the user's questions or proactively start conversations} based on contextual information about the current area of the robot, nearby exhibits, and areas that the user has visited.

To understand user perception of the remote experience via \system{}, we conducted a user study ($n=20$) where participants remotely visited a geology museum. Our analysis of interaction history logs and post-study interviews revealed user perceptions and interaction patterns related to dialogue-based \camera{place exploration in the real-world scenario. Drawing from insights from our field study, we identified future design opportunities and presented design implications for LLM-based narrative mobile robots.}

The contributions of our work are as follows: 
\begin{itemize}
    \item \textit{System contribution:} Designed and prototyped \system---an LLM-based narrative mobile robot designed to allow users to visit a geological museum remotely in real-time. \system{} autonomously navigated the environment, made suggestions for exhibits to visit, provided contextualized narratives on the exhibits, and responded to user questions.
    \item \textit{Empirical contribution:} Conducted field studies in a genuine geology museum to understand user perceptions of \system{} and the remote experience.
    \item \textit{Design implication:} Identified implications for the future design of LLM-enabled dialogue-based remote robotic technologies, highlighting robot narrative guidance to overcome uncertainty in physical navigation and information acquisition. 
\end{itemize}

\section{Related Work}

\subsection{Remote Experiences through Technologies}
Remote experiences refer to activities that allow users to access remote environments and gain new experiences through remote technologies. Users can utilize remote technologies to explore urban spaces \cite{heshmat2018geocaching}, experience nature \cite{tota2022telepresence}, or explore outer space \cite{bensch2024designing}. These experiences can be realized using various technology platforms, such as virtual reality (VR)~\cite{Cavallaro2021_VirtualTour, Teo2020_Exploring360Panoramas, Le2020_EnhancingVirtConf}, mobile videoconferencing~\cite{Lim2014_NomadCouchPotato}, livestreaming~\cite{Young2022_Telefest, Chen2022_LivstreamingFieldtrip}, telepresence robots~\cite{Neustaedter2016_BeamTelepresenceConference, Tsui2011_TelepresenceRobots, ng2015cloud, hu2025designing, hu2025multi}, \textit{etc}. \camera{Remote technology also provides a platform to design experiences by augmenting how users perceive live events with enhanced presentation and increasing access to location-specific activities, overcoming travel limitations to remote environments and experiences.} For example, \citet{Young2022_Telefest} presented a system for live-streaming mixed reality 360-degree videos for festival events such as concerts and \citet{deGreef2016_TeleTourist} invited tourists to have video calls to share their sightseeing views with remote people who want to travel but cannot leave their dwelling environment. Moreover, telepresence robots have been used to empower individuals with accessibility challenges~\cite{tsui2013designing, tsui2015accessible, hu2024homebound, barbareschi2023both} and participate in events and perform tasks as a proxy~\cite{Venolia2010_EmbodiedSocialProxy, rae2013influence, rae2014bodies, fitter2018evaluating}. For example, \citet{hu2024homebound} investigated the use of a conversational telepresence robot for homebound older adults to explore a botanical garden and a lakeside park remotely. Built on the prior work, we expanded the design space by investigating the use of robot guidance and narrative capabilities to support remote experiences.

 \begin{figure*}[!tb]
    \centering
    \includegraphics[width=\textwidth]{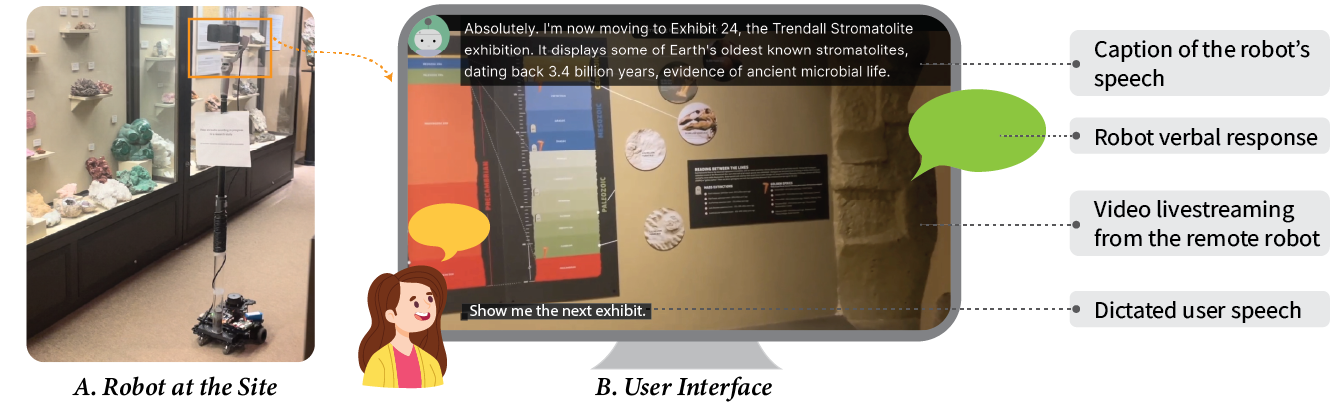}
    \caption{User interface displayed on a monitor. Remote users interact with the robot using dialogues and see the video livestreamed from the robot's camera.}
    \Description{}
    \label{fig: user interface}
\end{figure*}

\subsection{Robotic Guide for Remote Exploration}
Prior work has also honed focus on robots for the purpose of remote experiences \cite{burgard1999experiences, thrun1999minerva, shiomi2006interactive, trahanias2005tourbot} in different forms including mobile robots \cite{thrun1999minerva, burgard1999experiences} and wearable robotic devices \cite{manabe2020exploring, yazaki2023oh, yamada2022and}. For wearable robotic devices, the remote users' agency is executed through the local users who wear the devices. Users navigate in the remote place through a local user who carries the robot~\cite{manabe2020exploring}. These robots have been used for remote users to explore in the urban space and complete various tasks such geohunting and shopping~\cite{yazaki2023oh, yamada2022and, heshmat2018geocaching}. 

In addition to wearable devices, remote robots can be a standalone platform that allow users to take actions in the remote environment through the robot. These robots are remotely controlled by the user and provide information about the environment when requested \cite{ng2015cloud, burgard1999experiences, thrun1999minerva, trahanias2005tourbot}. One of the most studied platforms is the museum guide robot. For example, \citet{ng2015cloud} built a cloud robotics system that allow mobility impaired people to enjoy the museum through a telepresence robot. \citet{burgard1999experiences} and \citet{thrun1999minerva}~presented two generations of an autonomous and interactive tour-guide robot in a museum. The user navigated a web interface to interact with a robot where they can control the robot's movement, monitor the tour, get the information about the exhibit and join an online discussion forum. 
However, a common theme in the prior work that low-level piloting of remote robots, especially in crowded public space \cite{rae2017robotic, hu2024homebound}, can cause extra burden for remote users in the activities and took away from the experience \cite{Neustaedter2016_BeamTelepresenceConference, rae2014bodies, rae2017robotic, takayama2013presentation, yang2018shopping}. In our work, we explored the potential of dialogue-based robot interfaces powered by AI to provide narrative guidance for remote users, ease the burden of remote robot controls and deliver rich location-aware information through natural conversations.

\subsection{Conversational Agents for Remote Exploration}

Conversational virtual guides have been proven as an effective and engaging method for remote users to visit locations such as museums \cite{sylaiou2010exploring, sylaiou2020avatars} and shops \cite{Cavallaro2021_VirtualTour} and increase the sense of presence in the remote space \cite{sylaiou2010exploring}. These conversational agents often support virtual experiences through various media such as the website, projected screens, and augmented reality technologies in head mounted devices \cite{hammady2021framework, chu2024design} and mobile devices \cite{rojas2020application, wang2022survey, jacob2021collaborative}. Prior work has studied a variety of design factors of virtual guides including the avatar's storytelling voices \cite{sylaiou2020storytelling}, gaze induction \cite{aoyagi2024vr}, facial expressions ~\cite{Shamekhi2018_FaceValueGroupFacilitation} and explored different avatar roles in the visit \cite{sylaiou2020avatars}. While virtual guide presents an interactive and engaging way to convey new information for user's remote experiences, they lack the embodiment and the navigational capabilities in the remote space and only provide directional information for the user to go to different places. Built on the prior work of remote experience, our work studies the remote robot that not only are passively controlled by the user, but exhibit agency in decision making in the remote experience. We utilized a mobile conversational robot that is embodied in the remote environment, combines the capability of navigation and dialogue to support users' remote experiences.

\section{\oursys: a Narrative Mobile Robot for Remote Museum Experience} \label{sec_application}

As an instantiation of the interaction paradigm described in \S\ref{sec: intro}, we presented \oursys, a mobile robot that can engage in location-aware dialogues with a remote user and provide narrative guidance in visiting a geology museum. The dialogue content was derived from materials provided by the museum and the general flow of the conversation was grounded in observations from a tour experience on site and discussion with the assistant museum director.

\subsection{Artifact}

The remote user interacts with \system{} in the museum through a web interface. The robot has a smart phone mounted on top of a rod which is used to have a teleconferencing call\footnote{https://zoom.us/} with the remote user. The robot can navigate in the museum so the user can see the remote environment through the video call. The overall system set up and the user interface are illustrated in Figure \ref{fig: user interface}. 

\subsection{Interaction Design}
\subsubsection{Interaction Design}
Through \oursys's dialogue-based interface, the user can inquire about museum exhibits and control the robot's navigation in the museum. To create genuine robot narratives during tours, two authors observed one tour session, consisting of approximately 30 visitors at the museum. During the observation, the authors took field notes recording the types of questions asked and responses from the visitor. After the observation, the two authors had informal discussions with the assistant museum director, who has over 20 years of experience guiding tours, and identified the key factors in the tour: (1) answering visitor questions about the exhibit, (2) asking questions about the exhibit, (3) directing the point of interest in the museum, and (4) allowing for free exploration in the space. 
In addition to observing the guided tour, the assistant museum director also provided the authors with tour guide materials, such as a self-guided tour book and supplementary guide related to the exhibits.
Inspired by the provided materials, our tour observation, and related literature on tour practices~\cite{Best2012_MakingMuseumTours}, the research team conducted an informal preliminary study offsite at their research facility to simulate the robot conversing with a user and traversing from exhibit to exhibit.
Following this, the research team then deployed the robot in the museum after open hours to confirm the robot's ability to navigate and guide. 
At least one team member was onsite at the museum to observe and support the robot, and one team member joined online to verify the remote experience.

\paragraph{Robot Initiatives in Providing Narrative Guidance}
\camera{ 
The robot can proactively initiate dialogue with the user, introduce the exhibits that are nearby the user, ask them about their geology background, suggest the next exhibit to visit, and ask questions about the exhibits they have visited before. For example, the robot would proactively ask users, \textit{``Do you know which mineral in this case is the primary constituent of our teeth and bones?''} The narrative guidance of the robot is based on the tour guidebook information and includes a rich cultural and historical information about the exhibits. Below is an example where the robot introduced the local geology facts to the user: } 

\begin{quote}
\textbf{Robot}: ``Let's start with our first exhibit, which showcases a beautiful piece of red granite. Can you tell me what type of rock this red granite is?''

\textbf{User}: ``It is an igneous rock.''

\textbf{Robot}: ``That's correct. The red granite in this exhibit is indeed an igneous rock. It is formed through the cooling and solidification of magma. This particular specimen originated from the vicinity of Wausau, Wisconsin, and is estimated to be approximately 1.75 billion years old.''
\end{quote}

\paragraph{User Initiatives in Engaging with the Robot}
The user can ask the robot questions about the exhibits, instruct the robot to go to a specific exhibit, or control the robot by saying directional control commands, also known as low-level instructions, such as \textit{``turn left''} and \textit{''move forward.''}Below, is a sample dialogue when the user asked the robot about the current exhibit they are seeing:

\begin{quote}
\textbf{User}: \textit{``Tell me more about Galena.''} 

\textbf{Robot}: \textit{``Galena is formed in a wide range of hydrothermal environments and it is a component of some granites. } 
\end{quote}

The user can also provide high-level instructions, such as, \textit{``Go to the next exhibit.''}

\begin{figure*}[!tb]
    \centering
    \includegraphics[width=\textwidth]{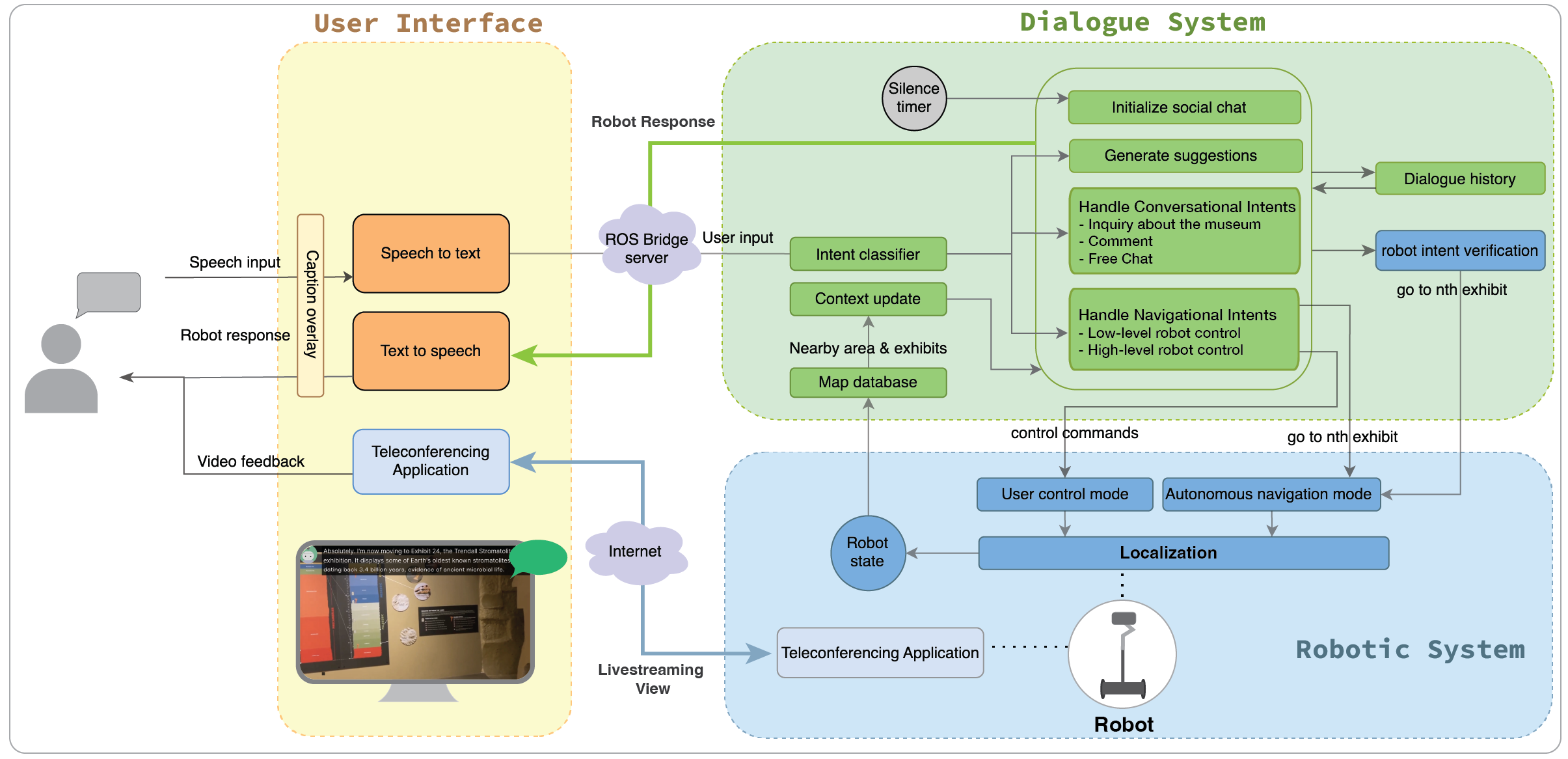}
    \caption{System architecture diagram for our user-robot interaction system, divided into three main modules: User Interface, Dialogue System, and Robotic System.}
    \Description{}
    \label{fig: system arch}
\end{figure*}

\subsection{Implementation}

\system{} is comprised of three main modules: user interface, dialogue system, and robotic system, as illustrated in Figure \ref{fig: system arch}. The user interface enables verbal interactions between the remote user and the robot in the museum. The dialogue system classifies the user's intent and generates responses in the context of the current exhibits. The dialogue system also interprets the user's control commands and sends navigation goals for the robot. The robot sends its real-time location to the dialogue system and navigates based on the user's instruction or following the curated routes. 

\subsubsection{User Interface}
A user interacts with the system through the dialogue-based interface hosted on a web browser. The user interface is implemented as a Chrome extension and overlays on top of a teleconferencing web page. The user interface uses a speech-to-text module\footnote{https://www.npmjs.com/package/react-speech-recognition} to convert the user's verbal input into text and sends it to the conversational robotic system. The robot's response is sent back to the user interface and converted to speech with a text-to-speech module\footnote{https://developer.chrome.com/docs/extensions/reference/api/tts}. ROS bridge\footnote{https://github.com/RobotWebTools/rosbridge\_suite} is used for the data transmission between the remote user and \system{}. 

\begin{figure*}[!tb]
    \centering
    \includegraphics[width=\textwidth]{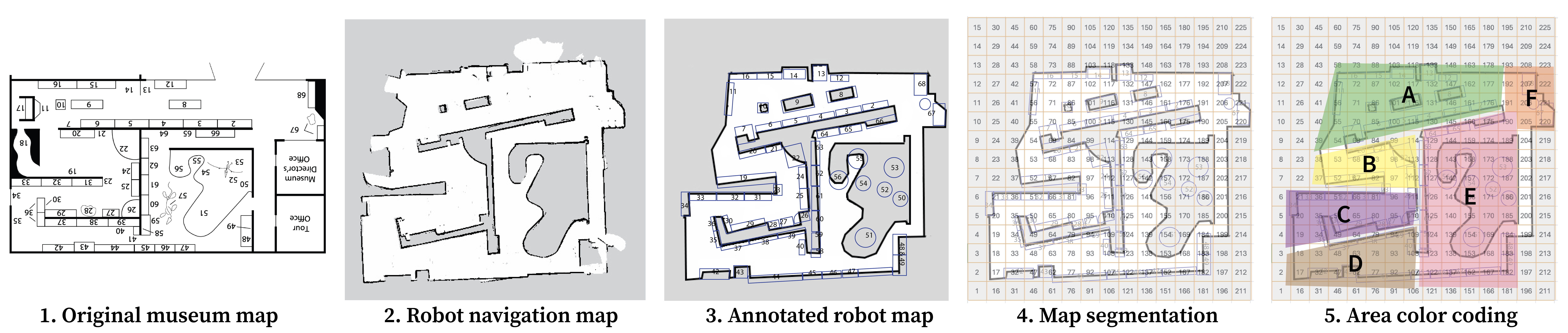}
    \caption{Creation, segmentation, and annotation of a map of the remote location for our system.}
    \Description{}
    \label{fig: map}
\end{figure*}

\subsubsection{Dataset Preparation}
The robot's narrative was generated from a museum dataset created by the research team. The dataset consolidated information from two tour guidebooks provided by the museum administrator and exhibit labels placed throughout the museum. Two members from the research team served as labelers for the dataset and categorized the data source based on the exhibit number. For each exhibit, the labelers identified content related to history (\textit{e.g.,} explanations of what the exhibit is and where it can be found in nature), activities (\textit{e.g.,} inviting a visitor to compare stones within the same exhibit case), and miscellaneous information (\textit{e.g.,} labels from diagrams and table content). In total, we prepared a dataset of all 68 exhibits in the museum. This dataset was further processed by GPT4 \footnote{https://openai.com/gpt-4} to generate two additional field for robot dialogue generation at a later stage: 1) introduction for each exhibit and 2) sample dialogues between a museum visitor and a tour guide about this exhibit. 

\subsubsection{Dialogue System}
The dialogue system is comprised of an intent classifier, a set of user intent handlers. The user input is first classified into intents and then the corresponding intent handler generates responses based on the user input and the current context information (\textit{i.e.}, current exhibits near the robot). The context information is retrieved from a map database annotated with exhibition information based on the current location of the robot. Both the intent classifier and handlers are powered by GPT-4\footnote{https://openai.com/gpt-4} where each of them has their own pre-prompts and parameters. All the metadata and parameters in prompting the language models are provided in the Appendix as the supplemental materials.

In handling the user's intent, the classifier first categorizes user inputs into conversational intents and navigational intents. Three conversational intents are \textit{inquiry about the museum}, \textit{comment}, and \textit{free chat}. Two navigational intents include \textit{low-level robot control} and \textit{high-level robot control}. The intent classifier was implemented with GPT-4 by providing classification examples and prompting the model to classify the new user input. 

The robot's prompt is either generated by the intent handler to respond to the user input or when the robot initiates a social chat after a 45-second period of silence. The 45-second time-lapse is selected based on experimental results from pilot tests to achieve a natural interaction between the user and the robot. Each intent handler has its own grounding information to prompt the LLM. 

\paragraph{Navigational Intent Handler}

After the initial intent classification, the low-level intents are mapped to the four control commands, \textit{i.e.}, turning left, turning right, moving forward, and moving backward. The high-level control intents are further processed to identify the next goal for the robot's navigation, \textit{i.e.}, the exhibit number, using GPT-4's function calling \footnote{https://platform.openai.com/docs/guides/function-calling}. The robot's navigational goal is detected either from the user's input or the robot's suggestion. For example, the user input \textit{``Can you show me exhibit 13?''} will detect the goal exhibit 13 for the robot to navigate to next.  If the user provided an open-ended instruction such as ``show me around,'' the robot will generate a destination suggestion based on the history of exhibits visited and the predefined order of the exhibits.

\paragraph{Conversational Intent Handler}
To handle the user input related to the three conversational intents, \textit{i.e.}, inquiry about the museum, comment, and free chat, the prompt sending to GPT-4 consists of the introduction and sample dialogues of the current exhibit, the area that the robot is in, and introductions of the nearby exhibits. Below is a prompt example when the user asks about Exhibit 4: 

\begin{quote}
The tour guide is having a dialogue with the visitor about the exhibit. 
Answer the visitor's question and ask the visitor about their experience about geology.\\

    The visitor is seeing or going to see the following Exhibit:\\
    This exhibit showcases a variety of yellow minerals, each possessing unique characteristics.
   One such mineral is Pyrite, often referred to as ``fool\'s gold'... 
    \\

     Here is a sample dialogue between the tour guide and the visitor:\\
      'Tour Guide: Can you identify the mineral in this case that\'s also known as ``fool\'s gold?'' \\
       Visitor: That would be Pyrite. \\
       Tour Guide: Correct! It often tricks miners into thinking they\'ve found real gold.\\

     This is the area the visitor is in:\\
    Introduction of the area......\\
     Introduction of each nearby exhibit......
\end{quote}


We reported the full prompt list for each intent handler and classifiers in the supplemental materials.

\subsubsection{Robotic System}
\paragraph{Hardware}
The mobile robot in the museum used TurtleBot3 Waffle\footnote{https://emanual.robotis.com/docs/en/platform/turtlebot3/overview/} as its mobile base. It was equipped with a Lidar sensor (LDS-01) and an IMU for SLAM and navigation. Robot Operating System 2 (ROS2)\footnote{https://docs.ros.org/en/humble/index.html}, a set of modular software libraries and tools for developing robot applications, was used as the middleware framework for integrating the robot’s sensors, navigation and control, and communication systems. A smartphone was mounted on top of the mobile base with an adjustable rod and a gimbal stabilizer, which was used for teleconferencing calls\footnote{https://zoom.us/} with local users remote users. 

\paragraph{Map Creation, Segmentation, and Annotation}
\label{map}
The robot navigation map was created using LiDAR-based SLAM. The environment was scanned using the robot's LiDAR sensor, and the resulting data was fused with information from the wheel odometry and IMU to enhance SLAM accuracy. The ROS2 package Cartographer was used for the SLAM process \footnote{https://github.com/cartographer-project/cartographer\_ros}. After the robot navigation map was created, the exhibits and their numbers from the original map were mapped to the robot navigation map. Additionally, the optimal pose (position and orientation) for viewing each exhibit was determined by manual exploration by two authors and recorded in the robot navigation map. The maps were then segmented into grids, and the grids are clustered into different general areas, and the nearby exhibits were grouped together according to the grids they are in. Therefore, given a specific grid the robot is in, the general area the robot is in and the exhibits near the robot can be known.

\paragraph{Robot Navigation}
The annotated robot navigation map, as detailed in \S \ref{map}, enabled us to guide our robot across the exhibits. To autonomously navigate the robot through the exhibits, the ROS2 package, Nav2\footnote{https://github.com/ros-planning/navigation2} was used and selected for its compatibility with our navigational needs. Two navigation modes were implemented: autonomous navigation mode and user control mode. The robot enters autonomous navigation mode when receiving the user's instructions to guide them or go to a specific exhibition such as ``go to the next one'' or ``go to exhibit 10''. The robot's path planner identifies the goal and creates a route for the robot to follow and navigate to the destination. When arriving at the destination, the robot can adjust to a pose that faces the exhibit so the exhibit is in view of the user. The robot's optimal poses for viewing the exhibit were stored for each exhibit beforehand. The robot can enter user control mode when receiving the user's instructions to move forward, backward, to the left, and the right. The robot follows the user's instructions to move accordingly.

\subsection{System Deployment}
We deployed the robotic system in a local geology museum. The geology museum has 68 exhibits in total across four areas: minerals and rocks, earth history, vertebrate fossils, dinosaur fossils, and meteorites. Due to the large space of the museum and the time limitation of experience, we constrained the robot's exploration to approximately half of the museum space, comprising 23 exhibits across three zones. To ensure we curated a coherent tour given our constraints, we consulted with an experienced geologist of the museum and identified a selection of 11 exhibits based on their prevalence in the local area of the museum and popular exhibits. Some examples of exhibits include local minerals and rocks and trace fossils.

\section{User Study}

To understand how \system{} can enable participants to explore remote spaces, we conducted a user study with 20 participants that experienced a tour session with \system{} in the geology museum. The following study was approved by the authors' Institutional Review Board.

\subsection{Participants}

We recruited 20 participants (P1--20; eight women, twelve men) aged 19--77 ($M=29$, $SD=13.57$) from the local university to participate in our user study. Recruited participants reported being fluent in English and not having any cognitive impairments as our system was not designed to support a broad range of cognitive abilities. Among the 20 participants, 11 have been to the geology museum before the study. A table with the demographic details of the participants and their familiarity with geology can be found in the supplementary materials.
Each participant was compensated with \$15 USD for their participation.

\subsection{Task and Procedure}

During each user study, at least two experimenters were present. 
After obtaining consent from the participant, the experimenter explained the purpose of the preliminary user study and task. 
The participants were tasked to engage in a tour session via \system{} to explore the museum for 20 minutes and visit at least five exhibits.
We selected five exhibits as the task goal to ensure that participants had adequate time to interact with \system{} at and in between exhibits. 
Each study was conducted individually, where the participant was co-located with an experimenter and 1--2 members of the research team were co-located with \system{} to ensure safe navigation throughout the museum (\textit{e.g.,} avoid collision with exhibits or visitors).


Before starting the tour session, all participants completed a tutorial, where they practiced speaking to the robot and familiarizing themselves with the conversational aspect and navigation of \system{} in conjunction with the remote experience. 
During the tutorial, the participants were also asked to practice using speaking with \system{} and also prompted to use direction phrases such as \textit{``Turn left,''} to ensure they were able to converse and navigate the museum via \system{}. 
After the experimenter confirmed the participant was familiar with the controls of \system{}, the participant was given 20 minutes to explore the museum. 
The experimenter prompted the participant five minutes before the end of the session.
Following the tour session, the experimenter then conducted a semi-structural interview with the participant, asking questions about their experience exploring the museum via \system{}, and concluded the session by administering a questionnaire. All interview questions and questionnaires are available in the supplementary materials.


\subsection{Data Collection and Analysis}
All tour sessions were audio- and video-recorded, and interview sessions were audio-recorded for further analysis.
As the museum was located in a public space, we placed signs on the body of \system{} and at the entrance of the museum to indicate that a recording was ongoing for research.
During the tour session, the system logged the dialogue data between the participant and \system{}.

\paragraph{Interaction History}
We reviewed the interaction log and video recordings to confirm the participants' interaction history with the conversational robot. We first characterize their conversation by noting the number of words used by the participant when conversing with the robot and the participants' responses to the robot's suggestions. 

To further investigate user speech patterns, we characterized user speeches into museum inquiry and robot control commands. Robot control commands were further categorized into low- and high-level instructions for the robot. Low-level instructions include directional control commands such as \textit{``move forward''} and \textit{``stop.''} High-level instructions include speeches directing to a navigational goal such as \textit{``Go to the next exhibit,''} \textit{``Show me around,''} and \textit{``Can you take me to exhibit four?''} We also characterized the politeness of the language used by participants based on a subset of categories, polite and direct (also known as bald on-record), from the Politeness Theory \cite{brown1987politeness, hu2022polite}. We refer to polite tones as instances where the participant softened their language by using positive or negative politeness strategies, \textit{e.g.}, adding linguistic markers \textit{``could you,''} \textit{``can you,''} \textit{etc}. in their utterance (e.g., \textit{``could you go to the next exhibit?''}). We refer to direct, or bald on-record, tones as instances where the participant directly conveyed their thoughts without using politeness strategies (\textit{e.g.}, \textit{``go to the next exhibit''} or \textit{``next exhibit''}).

Two members from the research team performed the coding task for the interaction history. One coder first independently coded the interaction history and the second coder verified the coding results by going through all the dataset.

\paragraph{Video Analysis}
We analyzed the screen recordings of the user interface with has views from the robot to understand bystanders' reactions to the robot. We first utilized a computer vision algorithm for human detection\footnote{YOLOv8: https://docs.ultralytics.com/models/yolov8/} to identify video clips where humans were present in the videos. One member from the research team reviewed the recordings and the human detection results, made modifications for any incorrect detections and added missed detections. Next, two coders performed the video analysis using a video annotation tool ELAN\footnote{ELAN: https://archive.mpi.nl/tla/elan} following the same coding process as the interaction history data described above.

\paragraph{Interview}
Furthermore, we conducted a thematic analysis of the interview transcripts.
Using an inductive (data-driven) approach, the first two authors from the research team independently coded all the interview data and identified themes that emerged from the dataset, such as the participants' impressions of the overall touring experience, their feedback on the conversational experience with the robot, suggestions on how to design the robot to enhance their remote touring experience, etc.
After agreeing upon an initial codebook, the two authors then coded the remaining interviews and derived themes through affinity diagramming. The themes were then reviewed with the other authors and revised into categories presented below.

\begin{figure*}[!tb]
    \centering
    \includegraphics[width=\textwidth]{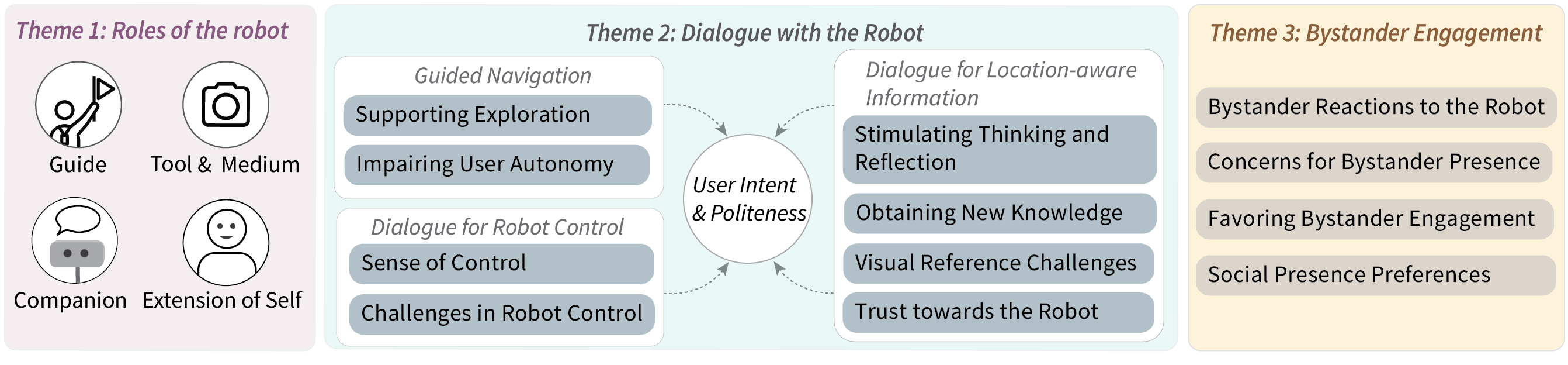}
    \caption{The themes that emerged in our data analysis, including the different roles that participants perceived the robot to play \textit{(left)}, the patterns in which they had dialogue with the robot \textit{(middle)} and user preferences for bystander engagement \textit{(right)}.}
    \Description{}
    \label{fig: theme}
\end{figure*}

\section{Findings} \label{sec:findings}
Below, we first report the quantitative results of the interaction history analysis, and then report our observations and qualitative findings from the user study. For the qualitative findings, we first present the perceived dynamic roles of the \system{}. Next, we report the user interaction patterns and feedback for the dialogue-based robot interface and the robot's narrative guidance. Lastly, we report findings related user feedback for encountering bystanders. We note that three participants (P2, 4, 13) experienced only 10 minutes of the tour because of technical issues such as system malfunctions and loss of Internet connection. We also note issues of the robot planner with two participants (P6, 7) which caused the robot made duplicate recommendations. While these incidents affected the usability of the system, we argue that the insights from these five participants are still valid and thus we included their experience.

 \subsection{Quantitative Results} \label{sec:quant res}
Paired sample t-test was performed for the following three comparisons: 1) times of user acceptance and that of rejections to robot suggestions, 2) times of museum inquiry speeches and that of robot control commands, and 3) times of high-level robot instructions and times of low-level robot instructions. Results showed that participants had significantly more acceptance ($M=2.90\ times$, $STD=2.32\ times$) than rejections ($M=0.9\ times$, $STD=0.99\ times$) in responding to the robot suggestions [$t(19) = 4.06, p = 0.0007$]. However, no statistically significant result was found for the difference between the usage of high-level instructions ($M=7.90\ times$, $STD=4.17\ times$) and low-level instructions ($M=7.35\ times$, $STD=7.26\ times$) when controlling the robot [$t(19) = 0.32, p = 0.75$]. Also the result was not significantly different between the amount of museum inquiry speeches ($M=14.65\ times$, $STD=9.41\ times$) and robot control commands ($M=15.50\ times$, $STD=9.36\ times$) [$t(19) = -0.25, p = 0.80$]. The quantitative results indicated that participants tended to accept the robot's suggestions than rejecting them.

\subsection{Perceived Roles and Identities of the Robot}

Participants shared their perspectives on the role the robot played during the remote tour, \textit{i.e.}, a guide, a tool and medium, a companion, and a representation of self in the remote location.

\subsubsection{\textbf{As a guide}}

Participants described the robot as a tour guide (P1, 7--11, 14, 18) and an instructor (P2, 3, 17). The robot's guidance supported the remote experience for participants who did not have knowledge of geology or the museum (P3, 8, 9, 11, 14). For example, P14 commented: \textit{``I think it was pretty helpful, because even I didn't know any of the exhibits beforehand. Once again, acting as a guide kind of really helped and made my experience much better, definitely.''} Participants commented that the robot was ``\textit{knowledgeable}'' (P3, 8, 11, 14, 18), ``\textit{intelligent}'' (P11), ``\textit{professional}'' (P1, 8, 11) and ``\textit{informative}'' (P8). One participant (P1) highlighted the convenience to ask the robot questions compared to a human tour guide: \textit{``I would feel that I'm bothering the tour guide, he might feel like I'm dumb... With a robot, I don't have to worry about that.''} However, several participants (P4, 6, 7, 13) commented that the robot's knowledge is generic and basic, in particular for participants (P4, 6, 7) who have had professional experience in geology.

\subsubsection{\textbf{As a tool and medium}}
Participants (P1, 2, 4, 10, 11, 13, 15, 16, 19) noted that the robot acted as a tool that enabled them to get new information and navigate through the museum. They viewed the robot as a ``\textit{learning tool}'' (P4) and ``\textit{Wikipedia}'' (P11) for obtaining information, and as a medium to connect to a real human guide if the robot doesn't have the knowledge (P5, 15). As P15 commented, \textit{``When the robot cannot understand [the exhibit], there's a person to help.''} (P15). As a navigation tool, participants used varying ways to instruct the robot to move. They have used a mixture of low-level control commands (\textit{e.g.,} ``\textit{Move forward}'') and high-level instructions such as ``\textit{go to the next exhibit}.'' 


\subsubsection{\textbf{As a Companion}}
Two participants (P12, 20) described the robot as their companion in the museum visit. One participant (P12) even preferred the robot's company than the company from a friend because the robot was \textit{``loyal''} and would not have disagreements with her on where to go next. As she reported, \textit{``He [The robot] will always go to the place that I want to and he's quiet and he was understandable.''} (P12). Another participant (P20) emphasized the \textit{``social aspect''} of having dialogue with the robot companion and described its role as \textit{``trying to engage the interaction between the user and the actual exhibits''} (P20).

\subsubsection{\textbf{As a representation of self}} Four participants (P1, 15, 16, 18) saw the robot as an extension of themselves. They abstractly described the robot as a \textit{``proxy''} (P16) and as their ``\textit{eyes}'' (P1, 15, 16) in the museum. For example, P15 commented, \textit{``It's like my remote eyes so I can, like, see it through the robot, see everything through the robot.''} Another participant (P18) saw the robot both as a guide and a proxy, commenting, \textit{``It feels like part of the guide and part of me sort of trying to navigate in a robot body in this.''}

\subsection{Dialogue with the Robot}
\subsubsection{Dialogue for Guided Navigation}
While exploring the museum, the robot made suggestions on what to see next, navigated to the destination and provided narrative guidance. On the one hand, the robot supported the user autonomy by helping them overcome uncertainties of not knowing the layout of the museum or not having knowledge of the exhibits during the visit. On the other hand, the robot's suggestions impaired user autonomy because of the lack of user confirmation and limited options provided in the suggestions.

\paragraph{Robot Suggestions Supporting Exploration} 
Most participants (P1--4, 6--8, 11, 12, 14, 18) appreciated the robot's suggestions. Six participants (P2, 8, 11, 12, 14, 18) mentioned that they did not have knowledge of the museum and described how the robot can ``\textit{direct}'' (P11) and ``\textit{lead}'' (P8) them in the navigation. For example, P8 appreciated the robot's support for finding new exhibits to visit: ``\textit{I really love this, this function, because again, I literally knew nothing about the whole museum and I don't know where we should go to find more like new, new rocks and fossil}'' (P8). Similarly, P12 described the robot as \textit{``a professional guy,''} commenting, \textit{``I think tak[ing] his or her opinion is also good way to explore the room.''} (P12). P13 also appreciated the convenience of having robot provide guidance: \textit{``You don't have to overthink about like which one is the best, maybe the best rock or and you can just rely on robot and with every command.''} (P13).

\paragraph{Robot Suggestions Impairing User Autonomy}
Alternatively, participants reported factors that can impair their autonomy when the robot made suggestions, including the lack of user confirmation (P4, 13), lack of contextual information (P15), and limited user options (P9, 10, 19). For example, P4 shared how the robot directed him to another exhibit when he \textit{``wasn't done at a certain exhibit.'' }(P4) This participant further suggested that the robot could ask \textit{``Would you like to move on to this''} instead of saying \textit{``Let's go do this.''} (P4). Another participant (P15) suggested that the robot could provide more contextual information when making the suggestions: \textit{``If the robot could tell me which I have visited and which I can explore afterwards, it would be good.''} (P15). Instead of being guided by the robot, two participants (P16, 19) preferred to control the robot on their own and ``go more freely'' (P19). P16 mentioned that she typically would only focus on things she was interested in during tours. However, the robot's guidance prevented her from staying on the exhibits of her interests. As P16 commented, \textit{``When I wanted to get closer to the exhibit, and then it just moved on to another exhibit. So that's what I would not really do, like explore everything and see what I am not really interested in.''}

\subsubsection{Dialogue for Robot Control}
In addition to follow the robot's guidance, users can instruct the robot to navigate in the museum. Participants reported various levels of sense of control and also challenges in controlling the robot through verbal interactions. 

\paragraph{Sense of Control} 

Seven participants (P1, 3, 4, 7, 11, 14, 17) highlighted the sense of autonomy with the robot following their commands. Participants shared how the robot followed both high-level instructions and low-level commands, saying that \textit{``I asked [it] to go to a specific, specific exhibit. And I went there. And I asked it like, about if you could show me fossils, and it went to fossils. So that was good. And then I could ask you to turn, and it turn for a little bit.''} (P4) and \textit{``All the commands that I gave it, it was following pretty well.''} (P14). Another participant (P3) shared that the experience through the robot was close to being physically in the museum, saying that \textit{``I feel like overall it was kind of tailored to how I would navigate that museum.''} (P3). On the other hand, eight participants (P5, 8, 9, 10, 12, 13, 15, 20) reported a mixed sense of control and five participants (P2, 6, 16, 18, 19) reported low sense of control due to verbal control challenges.

\paragraph{Challenges in Robot Control}
\camera{Participants reported a series of challenges in controlling the robot through the verbal interaction. These challenges included turn-taking issues, user intention detection failures and environmental noises.} First, participants (P2, 18, 19) pointed out turn-taking issues that they cannot interrupt the robot when it was speaking and they wished to control the robot during its speech. For example, P2 commented: \textit{``When he gives you the feedback, he cannot listen to the other instructions that you give to him.''} To overcome this challenge, P19 suggested to use mouse and arrow keys to direct the robot on top of the voice interaction. \camera{In addition, participants reported the robot's failure in detecting their intentions for fine-controlled movement.} They complained that the robot could not break down \textit{``complex instructions''}(P19), was inflexible for fine-controlled movement (P9, 19), and misunderstood their commands in adjusting the distance between the robot and the exhibits (P13, 16). P9 shared the difficulty they experienced with low-level details while verbally instructing the robot to make a turn: ``\textit{It's probably like a 15-degree turn, rather than a full turn... I feel like the freedom of moving around, it's kind of hard to navigate just by voice.}'' \camera{Participants (P13, 18) also reported issues in voice recognition from environmental noises, such as background sounds and bystanders' speech. P13 reported the confusion when the robot responded to a bystander, \textit{``I remember there's was a time that the robot posted, continuously posted four or five response at the same time, that really confused me.''} Another participant (P18) shared their preference to adjust the volume of the background dynamically: \textit{``You could dim down the [background] voices... when the robot is explaining, and then just go back to to regular voices and such, making it like you're in a museum.''}}

\begin{figure*}[!tb]
    \centering
    \includegraphics[width=\textwidth]{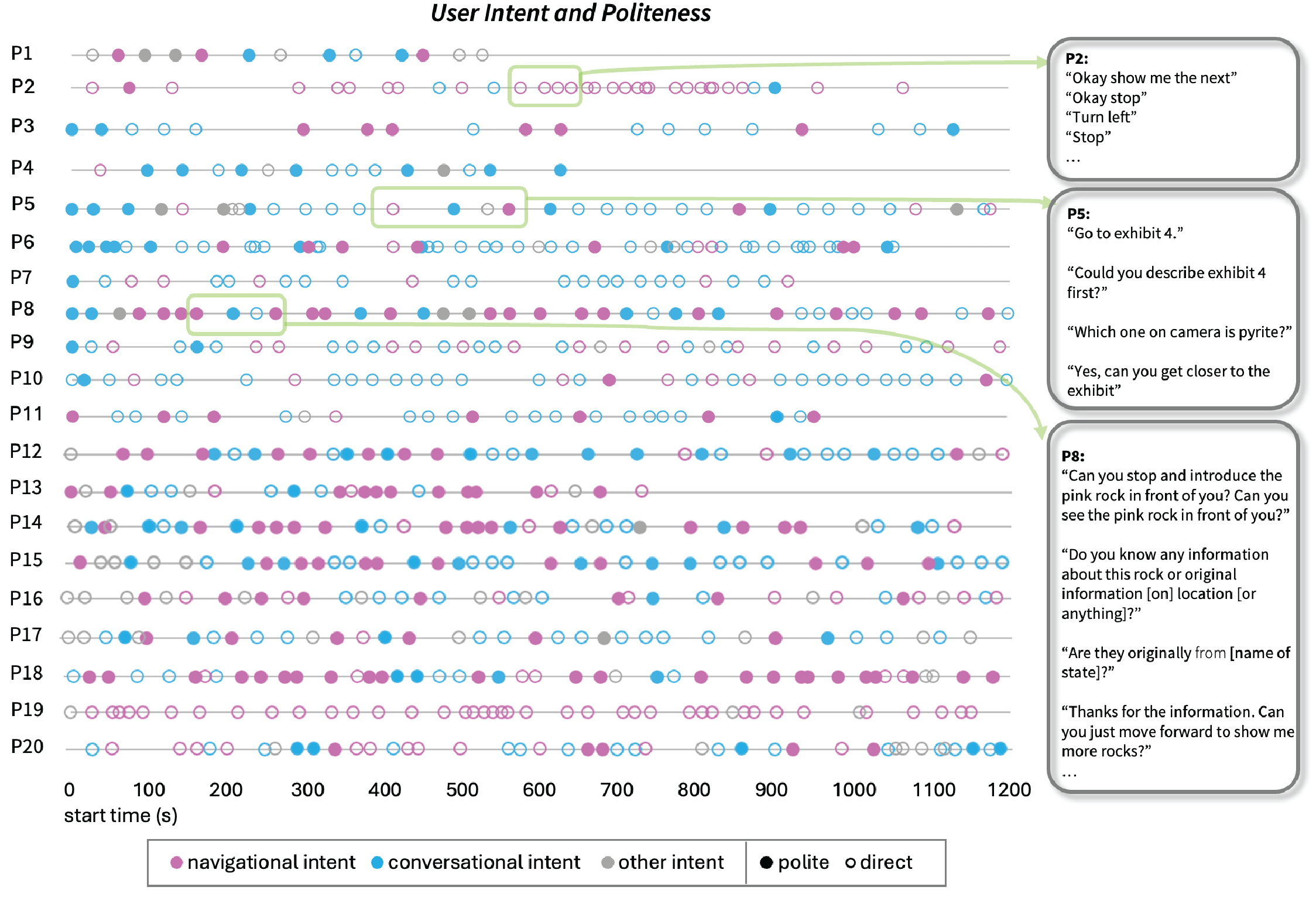}
    \caption{We mapped user speeches based on the types of intents and the polite or direct language used by the participant on the timeline to visualize their interaction.}
    \Description{}
    \label{fig:visualization}
\end{figure*}

\subsubsection{Dialogue for Location-aware Narratives}

We observed patterns in how participants interacted with the robot to obtain location-based information in the museum exploration. Participants shared that the robot's narratives stimulated their thinking and helped them gain new knowledge. In the meantime, participants reported the robot's limitation in providing visual references during its narratives. We also collected participants' perspectives on trust towards the robot and identified challenges for the user to detect fake information due to the lack of relevant knowledge.

\paragraph{Stimulating Thinking and Reflection}
The robot proactively asked questions from time to time in the remote tour and participants (P1, 4, 6, 15) shared how the robot's questions stimulated their thinking, ``\textit{lure[d]}'' their curiosity, and got them to ``\textit{think a little bit}'' (P4). As P1 described: ``\textit{They sort of challenge you to think more about the exhibition. And then, and then it sort of lures your curiosity to learn more about the exhibition.}'' P6 also noted that the nature of the speed of the conversation made them ponder more about exhibits, sharing that ``\textit{It made me slow down and think about what I wanted... When you're... forced to slow down and focus on something like that... it becomes a deeper experience.}''(P6). P15 also shared that the robot's questions such as \textit{``Do you know why this kind of rock?''} raised his interest, commenting that, \textit{``It will raise my interest in exploring, because otherwise it's like a normal rock.''} In addition, P6 mentioned that the the conversation facilitated recalling prior geology knowledge: ``\textit{The information that I have in my head, I forget it, until you remember it again, or reminded of it.}''

\paragraph{Obtaining New Knowledge}
Participants (P1, 2, 10, 11, 12, 15, 20) regarded the remote touring experience as an opportunity to obtain new knowledge. In particular, participants compared their in-person experience at the museum and the remote visit with the robot and shared that the robot helped them learn things that they would not have known otherwise. For example, P2 emphasized his experience with the robot: ``\textit{The good thing is like [robot] can give you some information that... you would never know if you just go by yourself here.}'' (P2). P11 also shared how the dialogue with the robot eased her learning process that the robot's narratives helped her obtain information that she often ``\textit{don't have the patience to look at}'' if she visited the museum in person. In addition to the robot's narratives, two participants (P1, 2) noted that the real-time Q\&A allowed them to more flexibly inquire about the museum in comparison with searching on the phone or listening to a pre-recorded guide. 

\paragraph{Visual Reference Challenges}
When the robot was explaining multiple rocks and minerals in one display, multiple participants (P1, 2, 4, 5, 7, 9, 11, 12, 17, 19) reported challenges in referencing the exact item in the exhibit. As P5 commented: ``\textit{It cannot, like to help me to identify which kind of mineral it is referring to when it says a name.}'' Participants suggested various ways to enhance the visual reference when the robot provided explanations, including the use of gestures and visual augmentations. One participant (P12) wished that the robot had arms and could use gestures to point to the object during the dialogue. Three participants (P7, 19, 20) suggested that the robot could use the camera to direct the user's attention, such as ``center[ing]'' (P20), ``zoom[ing] in'' (P19) and ``focusing on a particular specimen'' (P7). One participant (P5) also suggested to augment the visual experience by superimposing complementary images and description of the exhibit. As P5 described, ``\textit{If the user can like point to a label, we can enlarge the label so that the user can see like the label and the mineral in details.}'' (P5).

\paragraph{Trust towards the Robot}
When asked about if they trust the robot, majority of our participants (P2, 3, 5, 8--11, 13, 14, 18, 20) reported that they trusted the robot's narratives and thought it was ``\textit{accurate}'' (P8, 9, 20) and ``\textit{correct}'' (P2, 9, 13), and \textit{``answered my questions pretty concisely''} (P14). Some participants (P13, 18) trusted the robot's narratives because they did not have knowledge in geology, commenting that, \textit{``it feels like right''} (P18). Although some participants (P4--7, 13, 20) pointed out that the information was not completely correct, they still found the robot's explanation satisfying, saying that ``\textit{I didn't see any glaring issue}'' (P4).  In addition, we have observed instances where the robot came up with fake information from hallucination, however, participants couldn't detect the fake information. For example, two participants (P1, 8) thought the robot correctly detected the colored rock they pointed to and answered their questions even though the robot had made up the answer.

\subsubsection{Dialogue Intent and Politeness}


After verifying the utterances from participants' dialogue with the robot, we identified patterns with varying politeness (\textit{i.e.}, direct or polite) and type of speech intents (\textit{i.e.}, navigational,  conversational, or other intents) to understand how users interact with the robot verbally. For the use of politeness strategies, most participants used a mixture of direct and polite languages, while two participants (P19, 20) primarily used direct tones and four participants (P8, 12, 14, 18) mainly used polite tones. In terms of the speech intents, most participants exhibited a mixture of conversational intents and navigational intents, while three participants (P2, 8, 19) primarily interacted with the robot with navigational intents and another two (P5, 6) with conversational intents. In Figure~\ref{fig:visualization}, we visualize the interaction for all participants, indicating polite or direct tones and the three types of intents, and present a detailed breakdown of select snippets of observed interaction. Below, we present a selection of instances when the participant interacted with the robot.

We observed instances where the participants responded to the robot with a direct tone when navigating around the museum and a polite tone when have a conversation. For example, P5 used concise languages when navigating around the museum, \textit{e.g.},\textit{``go to exhibit four''} and polite languages when inquiring about the museum, \textit{e.g.}, \textit{``could you describe exhibit four first?''}
Conversely, we observed participants using polite terms when instructing the robot on how to navigate in the museum. P11 used a polite tone when interacting with the robot for navigation: \textit{``Can you turn right [to] go to the next exhibit?''} and direct tone when inquiring about details in the museum: \textit{``How is the mineral formed?''} Politeness also changed as the interaction evolved as one participant (P5) initially used polite language, such as \textit{``Can you get closer to the exhibit?''} and later shifted to direct tones, \textit{``Okay, go to the next exhibit.''}

\subsection{Bystander Engagement}
Although the robotic system was not intended for social interactions, bystanders in the museum were seen by the remote participants through the robot. Below we report bystander reactions to the robot and then present user feedback for bystander presence and engagement. 

\subsubsection{Bystander Reactions to the Robot}
Since there was a signs on the robot indicating that recording is in progress for a research study, most bystanders who noticed the robot did not actively interact with the robot and walked away after the encountering the robot. Nevertheless, we observed unexpected ways that several children visitors interacted with the robot. In one session (P8), a group of children approached the robot and tried to dance in front of the camera on top of the robot. In another session (P14), two kids were first scared by the robot and hidden behind their guardian. After observing the robot for a while, they ran away from the robot while saying \textit{``Hi robot''}. Another kid tried to follow the robot and block it (P16). In addition, we also observed visitors watching the robot out of curiosity in several sessions (P2, 4, 6, 8).

\subsubsection{Favoring Bystander Engagement}
Encountering bystanders was favorable for seven participants (P10, 12, 15, 16, 18--20), enhanced their engagement in the remote tour and increased the playfulness of the experience. Participants shared that they were \textit{``more interested in people than geology.''}( P16), thought it was \textit{``fun''} (P18) to see bystanders' reactions to the robot, and felt ``\textit{socially engaged}''(P10) in a shared space with the people physically present in the museum. If engaging with bystanders, participants reported that they preferred to have small talk interactions (P12, 15, 16). For example, P16 wanted to hear about people's thoughts on exhibits they visited together and if \textit{``we felt completely different''} on the same things. Another participant (P15) wanted to meet people with shared background, sharing that \textit{``I just felt it's interesting. Like, `Hi, I'm the robot,' And maybe I can interact with them. Like, maybe if I see my friends here, because I, I am a [College Name] student.''} (P15).

\subsubsection{Concerns for Bystander Engagement}
On the other hand, encountering bystanders was not always desirable (P8, 13, 14, 17), in particular for participants who wanted to focus on the exhibits without distraction (P13, 14). Participants shared concerns about their social presence (P8), and did not want to bother other people in the museum (P8, 14). As P14 commented, \textit{``I guess the whole point of this was to focus on the museum.''} In addition, bystanders can block the camera as they walked in front of the robot and affected the view of remote users. For example, P13 found bystanders \textit{``annoying''} when his view was blocked, commenting that \textit{``The bystander may just move like across the [exhibit], what do you see right now.''} Furthermore, P8 expressed her concerns about her social presence through the robot and was worried bystanders would hear her non-native speaking or detect her lack of knowledge in the geology terminology. As she shared: \textit{``Because I'm not familiar with all the terminology and I'm not a native speaker, and I was worried that if they could hear me they may feel this was just a non-native speaker saying some wrong terminology or something. I was also nervous about this... I'd rather they don't hear me.''} (P8).

\subsubsection{Social Presence Preferences}
While the user's face wasn't presented on the robot, participants noted their thoughts on their self-presentation during their remote museum experience. Some participants (P12, 16) wished to be seen on the display of the robot and some participants (P8, 15, 18, 20) reported various reasons that they did not want to show their face on the robot. P15 preferred to hide the face to trick passersby for fun. As he commented, \textit{``I think if without my face, it will be more interesting, because the people don't know why.''} (P15). P20 preferred to not show the face due to the loss of social cues on the robot. She reported the concern for being misunderstood: \textit{``Once I'm behind a screen, I kind of lose a lot of social cues, and so might be doing things that I don't realize that I want to do.''} In addition, P18 suggested the use of avatars on the robot to preserve user privacy that users can choose to change their face into an \textit{``emoji''} (P18). He further explained this avatar feature can be helpful for introverted users: \textit{``Especially introverted people, but they still want to have a chance to talk. They could use that avatar instead of, like this directly displaying their faces.''}

\section{Discussion}

\camera{LLM-based agents are increasingly investigated for environmental exploration, however, the majority of the existing work focused on agents in virtual environments such as in mixed reality and web-based interfaces \cite{zhang2025effects, wang2024llmtour, rist2024explore}. In this work, we integrated the LLM-based narrative guidance into a physical mobile robot for remote space exploration in real-time. Through the field study in a geology museum, we identified how the robot augmented the remote user's visiting experience with rich geological, cultural and historical information in its narratives. Our findings also highlighted the social dynamics among the remote user, the robot, and bystanders in the environment and revealed both the opportunities and challenges of bystanders' presence in the remote experience.} Below, we present the implications of our work, study limitations, and future applications.

\subsection{Design Implications}
\camera{
Overall, participants shared their perception of the robot as a guide, a tool and medium, a companion or an extension of themselves during the remote tour experience. Participants employed different politeness strategies and even \textit{``code-switch''} between conversing with the robot with different roles. As a result, future robot design needs to consider the user's preferred roles of the robot for in the remote experiences \cite{mutlu2021virtual}. Designing a companion remote robot may incorporate social behaviors of the robot \cite{sullivan2025primal, hu2021creative}, whereas the design the robot as a tool should prioritize the efficiency of conveying the information to the remote users. If the user perceives the robot as a proxy of themselves, it is essential to consider the remote user's self-representation on the robot \cite{fitter2021wigging, yuji2024identity}. Below we report four major design factors of the narrative mobile robots for supporting the remote place exploration.}

\subsubsection{Designing for LLM-based Narrative Guidance}

The robot's LLM-based narrative guidance can help the user gain new information in a remote environment and access areas that they may not know ahead of time. The lack of environmental awareness is a known key barrier to a user's control of the robot in the remote environment \cite{heshmat2018geocaching, hu2024homebound}. As a result, the robot's guidance can support users to navigate in an unknown and new environment \cite{kuribayashi2025wanderguide, kuribayashi2023pathfinder}. \camera{As an initial step to understand the location-aware guidance generated by LLM, \system{} utilizes a 2D semantic map to generate the robot's dialogue with rich contextual information. Future work can utilize 3D-based information for the robot to reason the spacial relationships of objects in the space to better guide remote users. For instance, the user may describe a relative location of the object and ask for the robot's explanation (\textit{e.g.}, \textit{``What is the fossil on the right?''}). In this case, the robot needs to identify what the user refers to and provide explanations accordingly, which can be particularly challenging when the physical environment is complex.}

\camera{Due to uncertainties of the open dialogue between the user and the robot, we observed that the robot provided false information in some circumstances from hallucination \cite{achiam2023gpt}. Such hallucinations can be especially harmful for users who are not knowledgeable about the environment, as we observed that participants still expressed their trust towards the robot when hallucinations occurred.} This points to the need for the robot to be trustworthy and credible as an individual agent interacting with the remote user. Existing work has investigated strategies to repair or recover resiliently from failures or unexpected events to maintain trust and acceptance between the user and the robot \cite{lee2024rex}. Future work can explore incorporating such strategies for continuous operations in dynamic, unexpected settings.

\subsubsection{Designing for LLM-based Robot Control}

\system{} can both guide the user to navigate to the next destination and be controlled by the user. Our findings revealed the potential risk of impairing user autonomy in the robot guidance and suggested that the robot can provide more options for the user to choose from, and ask the user for confirmations to protect user autonomy. These findings offer design implications for future robot guidance design that adapts to users' preferences for levels of robot initiatives, \textit{e.g.}, how much control users should have over the robot's decisions and how users should be involved in the decision-making process of the robot \cite{praveena2023periscope, kamikubo2025beyond, hu2021creative}. Our participants also reported challenges in using dialogue to control the motion of the robot due to turn-taking inefficiency and lack of customized command detections. \camera{Prior work has investigated the use of LLMs for robot task planning and code generation \cite{vemprala2024gptforrobot, arenas2024promptrobot}, however, it is a known challenge to synthesize reliable and safe robot programs from LLMs \cite{yang2024safe}. To mitigate these challenges, prior work has integrated program verification into LLM-based end user interface for robot programming~\cite{karli2024alchemist} and developed design guidelines to support customization, consistency and multi-step instructions in LLM-based robot interfaces \cite{padmanabha2024voicepilot}. Mobile robots used for place exploration often operate in dynamic environments where they can encounter bystanders and unknown objects. Future work can integrate formal methods to verify robot programs \cite{porfirio2018veri} generated from LLM-based interfaces considering dynamic contextual information and environmental constraints.}

\subsubsection{Designing for Visual Interfaces for Remote Robot Exploration}
While verbal interactions are the primary modality for \system{}, our findings pointed to design opportunities of the robot's visual interface to overcome of the limitation of the dialogue interface. Our participants expressed the desire for the robot to visually reference objects being described during its dialogue. Therefore, the visual interface can enhance the robot's capability to convey environmental information. For example, within the context of the museum, exhibits can be placed in a shared case, where many objects can be clustered together. In these environments, a system can use computer vision-based techniques to highlight or annotate the object in question and ensure that both the robot and user are referring to the same object. Alternatively, future work may explore other forms of visual augmentation, such as mixed reality (MR), to present images or other supplementary media (\textit{e.g.}, videos related to the exhibits, or web links to information about the exhibits) to complement live remote touring experiences. 

\subsubsection{Designing for the Presence of Bystanders}
Although our robotic system was not intended for social interactions and the user's face was not displayed on the robot, our study uncovered interesting findings related to the social participation preferences of remote users and local people's reactions to the robot. During our study, people in the museum approached the robot to engage with it in unique and sometimes unexpected ways (\textit{e.g.}, dancing in front of the robot), echoing findings from existing literature on deploying telepresence robots in public spaces \cite{heshmat2018geocaching, yang2018shopping, hu2025designing}. 

Different from the conventional use of telepresence robots to support social communications, we observed how encountering people through the robot can negatively impact the user experience and impair their interaction goals. Several participants saw visiting exhibits in the museum as the primary goal, and the interaction with other people was distracting for them while touring the museum. The presence of bystanders occasionally blocked the view from the robot, and users were not able to see the exhibits fully. Moreover, overhearing the conversations among bystanders could be disturbing for the remote visitors, and the environmental sounds can affect the robot's voice recognition for the dialogue-based interaction. These findings suggest that it is challenging to deploy the narrative mobile robot in crowded environments, and the robot design needs to minimize the negative effects from the crowd on both the visual and audio experiences. For example, the robot can dynamically adjust its height and view angle to avoid being blocked by people in front of it, or adjust the volume of the background sound based on the sound source to preserve the ambient live atmosphere of the museum while filtering out environmental noises. 

\subsection{Limitation \& Future Work}

Our study took an initial step to understand how a narrative mobile robot could be used for remote place exploration and uncovered users' emerging behaviors when interacting with this robot. Findings and design implications from our current work can be used to develop a more refined system and inform comparative studies to establish the extent of these design factors (\textit{e.g.}, robot being conversational or informational, levels of user control, alternative input modalities, types of visual augmentation, \textit{etc.}) can affect user experience as a next step. Future work could compare the use of an AI guide and a local human guide accompanying a robot to further investigate the remote experience. In addition, we also acknowledge the following limitations in our study. 

First, in order to circumvent technical issues associated with moving across large areas, the location of our study was conducted in an indoor museum with a selection of exhibits. However, the application of our findings are not limited to indoor remote place exploration. Future work could enhance the robot's navigational ability to adapt to different terrains and study the remote experiences in outdoor environments. Studies in in-the-wild settings can expand upon our findings to broader contexts, allowing an understanding of how robots should be designed to facilitate remote experiences in diverse, unstructured environments.

Second, given the prototype nature of the robotic system and uncertainties in the field deployment, five of our participants experienced unexpected incidents during the study as reported in \S\ref{sec:findings}. \camera{Several user studies were conducted in parallel with school tours and private groups of visitors in the museum. Noises and presence of the crowds interrupted the user's dialogue with the robot and the robot's navigation. We also observed hallucination from the LLM during the study. To lower the risk of hallucination, we tuned model parameters (\textit{e.g.}, temperature) to reduce the randomness of the generation and prompted the robot to only use the provided context. However, as participants asked the robot about information beyond its knowledge base, hallucination occasionally occurred.} Although these incidents had minimal impacts on our investigation of how participants perceive the robot's role and their attitudes towards the robot's narrative guidance, future work needs to improve the robustness of the robotic system for more extensive field testing. Future work can involve exploring how robots should respond, adapt, and recover in such scenarios, enabling continuous operations in unforeseen environments. 

Lastly, we acknowledged our limitation in the design procedures of the remote tour experience in that we only sought consulting from one experienced tour guide and designed the experience based on our observations of the museum tours. We suggest future work to consider involving various experts of different experiences and touring practices to ensure a wide variety of styles are incorporated into the design of the robot dialogue and guidance. 

\section{Conclusion}

In this work, we presented \system, an LLM-based narrative mobile robot that can guide people to explore a remote place. Using museum touring as an example scenario, we deployed \system{} in a geology museum and conducted user studies to understand user preferences for the dialogue-based remote robotic experience. Our findings highlighted various user orientations towards the robot, user preferences for the robot's guidance, and opportunities and challenges for using dialogue to interact with the robot. We also observed interesting bystander reactions to the robot, identified interaction barriers and benefits for involving bystanders in the remote experience. Our findings led to design implications on LLM-based robot interfaces for narrative guidance, control and visual interface augmentation. Future work can continue deepening the exploration in a broader context for the LLM-based remote robotic experiences and investigate factors to overcome the interaction barriers and enrich the remote user experiences. 

\begin{acks}
This work received support in part from a Google Award for Inclusion and NSF awards 1925043 and 2312354. Figure 1--2 modified images by freepik for its design. We would like to thank our participants for their time and participation in this research study. We thank University of Wisconsin--Madison Geology Museum for sharing tour guide materials and providing a site for us to run the robot. Lastly, we thank Caroline Zhang and Mrigya Kumar for their support in running user studies. 
\end{acks}

\balance
\bibliographystyle{ACM-Reference-Format}
\bibliography{references}

\end{document}